\documentclass[12pt]{article}
\usepackage{color}
\usepackage{amssymb}
\usepackage{graphicx}
\usepackage{epstopdf}
\usepackage[hang,small,bf]{caption}
\usepackage{hyperref}
\usepackage{amsmath}
\usepackage{latexsym}

\newcommand{\be}[3]{\begin{equation}  \label{#1#2#3}}
\newcommand{\ee}{\end{equation}}
\newcommand{\ba}{\begin{array}}
\newcommand{\ea}{\end{array}}
\newcommand{\bea}[3]{\begin{eqnarray}  \label{#1#2#3}}
\newcommand{\eea}{\end{eqnarray}}

\let\Large=\large
\let\large=\normalsize

\newcommand{\haken}{\mathbin{\hbox to 8pt{%
                 \vrule height0.4pt width7pt depth0pt
                 \kern-.4pt
                 \vrule height4pt width0.4pt depth0pt\hss}}}

\def\openone{\leavevmode\hbox{\small1\kern-3.8pt\normalsize1}}

\def\bo{{\raise.15ex\hbox{\large$\Box$}}}               
\def\face{{\raise.2ex\hbox{$\displaystyle \bigodot$}\mskip-2.2mu \llap {$\ddot
        \smile$}}}                                      

\def\leftrightarrowfill{$\mathsurround=0pt \mathord\leftarrow \mkern-6mu
        \cleaders\hbox{$\mkern-2mu \mathord- \mkern-2mu$}\hfill
        \mkern-6mu \mathord\rightarrow$}       
\def\dvec#1{\vbox{\ialign{##\crcr
        \leftrightarrowfill\crcr\noalign{\kern-1pt\nointerlineskip}
        $\hfil\displaystyle{#1}\hfil$\crcr}}}           

\def\beq{\begin{equation}}
\def\eeq{\end{equation}}

\def\beqx{\begin{displaymath}}
\def\eeqx{\end{displaymath}}

\def\beqa{\begin{eqnarray}}
\def\eeqa{\end{eqnarray}}

\begin{document}
\DeclareGraphicsExtensions{.jpg,.pdf,.mps,.png}
\begin{flushright}
\baselineskip=12pt
EFI-09-12 \\
ANL-HEP-PR-09-32
\end{flushright}

\begin{center}
\vglue 1.5cm

{\Large\bf Prospects for MSSM Higgs Searches at the Tevatron } \vglue 2.0cm {\Large Patrick Draper$^{a,c}$, Tao Liu$^{a}$, and Carlos E.M.
Wagner$^{a,b,c}$}
\vglue 1cm {
$^a$ Enrico Fermi Institute and
$^b$ Kavli Institute for Cosmological Physics, \\
University of Chicago, 5640 S. Ellis Ave., Chicago, IL
60637\\\vglue 0.2cm
$^c$ HEP Division, Argonne National Laboratory,
9700 Cass Ave., Argonne, IL 60439
}
\end{center}

\vglue 1.0cm
\begin{abstract}
We analyze the Tevatron reach for neutral Higgs bosons in the Minimal Supersymmetric Standard Model (MSSM),
using current exclusion limits on the Standard Model Higgs. We study four common benchmark
scenarios for the soft supersymmetry-breaking parameters of the MSSM, including cases
where the Higgs decays differ significantly from the Standard Model, and provide projections
for the improvements in luminosity and efficiency required for the Tevatron to probe sizeable
regions of the $(m_A, \tan\beta)$ plane.
\end{abstract}

\newpage

\section{Introduction}

The minimal supersymmetric Standard Model (MSSM) is an ultraviolet completion of the
Standard Model (SM) that provides a technical solution to the hierarchy problem, is consistent
with the unification of gauge couplings at high energies, and includes a natural dark matter
candidate~\cite{reviews}. The model includes two Higgs doublets, and for supersymmetric
particle masses smaller than about 1~TeV, contains in most of the parameter space a Higgs with SM-like couplings to the vector gauge bosons and a mass that is smaller than about 130~GeV~\cite{Haber:1990aw},\cite{Carena:1995bx},\cite{Heinemeyer:1998},\cite{Espinosa:2000},\cite{Carena:2000npb},\cite{Degrassi:2001},\cite{Martin:2003}. The MSSM also contains non-standard CP-even
and CP-odd neutral Higgs bosons, which present enhanced couplings to the down quarks and
SM charged leptons relative to the couplings of the SM Higgs. Probing the MSSM Higgs sector
therefore demands a combination of SM-like as well as non-standard Higgs boson searches.

Possible searches for SM-like Higgs bosons at the Tevatron colliders have been
analyzed in the literature~\cite{Stange:1993ya},\cite{Spira:1998wh},\cite{Berger:2003pd},\cite{Anastasiou:2008tj},\cite{deFlorian:2009hc},\cite{Dittmaier:2003ej},\cite{Dawson:2005vi},\cite{Carena:2000yx},\cite{Han:1998ma},\cite{Han:1998sp}.
Recently, the CDF and D$\emptyset$ collaborations released a combined analysis of the Tevatron search
for the Higgs boson in the Standard Model, resulting in the exclusion of the SM Higgs at
95\% C.L. in the mass range 160-170 GeV and improved bounds on the Higgs production cross section in
the lower mass range~\cite{Phenomena:2009pt}. The inclusion of multiple Higgs production and decay
channels in the analysis was crucial in obtaining the new limits, particularly in the low-mass region.

Some Higgs searches at the Tevatron, particularly in the
$\tau^+\tau^-$ channel,
have already sought Higgs bosons with non-SM-like gauge couplings directly~\cite{Balazs:1998nt},\cite{Baer:1998rg},\cite{Drees:1997sh},
\cite{Carena:1998gk},\cite{Belyaev:2002zz},\cite{Abulencia:2005kq},\cite{:2008hu}.
These limits are relevant for the search for non-standard Higgs bosons in the MSSM. In the following, we will consider the limits from these searches independently before combining them with the limits from the SM-like Higgs searches, and find a complementarity in the coverage of MSSM parameter space.

In this note, we quantify the improvements in luminosity and signal efficiency necessary for the
Tevatron to place exclusion limits on the Higgs sector of the MSSM. We begin in section 2 by reviewing
briefly the MSSM Higgs spectrum and the coupling strengths relative to the Standard Model.
Then, in section 3 we discuss the translation of Standard Model exclusion limits into limits on the MSSM.
We examine four benchmark scenarios in the MSSM parameter space and analyze the limit-setting
potential of the Tevatron according to a range of increases in luminosity and efficiencies
projected by the experimental collaborations. If the experiments can achieve these improvements,
we find that it will have significant exclusion coverage of the MSSM Higgs sector parameter space.
Moreover, even parameter regions previously considered inaccessible to the Tevatron due to a
suppression of the branching fraction of $h\rightarrow b\bar{b}$ appear accessible through the
$h\rightarrow W^+W^-$ channel. For convenience, all figures presented in this paper are available online at \url{http://home.uchicago.edu/~pdraper/MSSMHiggs/MSSMHiggs.html}. We summarize our conclusions in section 4.

\section{MSSM Higgs Spectrum and Couplings}
The Higgs sector of the MSSM consists of two complex scalar doublets $H_1$ and $H_2$,
which present tree-level couplings to only down-type quarks and up-type quarks respectively.
$CP$ is conserved in the tree-level scalar potential, so the neutral fields split into two
$CP$-even mass eigenstates $h$ and $H$ and one $CP$-odd mass eigenstate $A$. The
$CP$-even mass matrix at tree level is given in the ($H_1$,$H_2$) basis by
\begin{equation}
\mathcal{M}^2=\left(
  \begin{array}{cc}
    m_A^2\sin^2\beta + m_Z^2\cos^2\beta & -(m_A^2+m_Z^2)\sin\beta\cos\beta \\
    -(m_A^2+m_Z^2)\sin\beta\cos\beta & m_A^2\cos^2\beta + m_Z^2\sin^2\beta \\
  \end{array}
\right).
\end{equation}

The $CP$-odd mass $m_A$ and $\tan\beta$, the ratio of the vacuum expectation values of the
neutral components of the two Higgs doublets, are inputs of the tree
level mass matrix. $\mathcal{M}^2$ can be diagonalized by introducing a mixing angle $\alpha$:
\begin{equation}
\left(
  \begin{array}{c}
    h \\
    H \\
  \end{array}
\right)=\left(
          \begin{array}{cc}
            -\sin\alpha & \cos\alpha \\
            \cos\alpha & \sin\alpha \\
          \end{array}
        \right)\left(
                 \begin{array}{c}
                   H^0_1 \\
                   H^0_2 \\
                 \end{array}
               \right).
\end{equation}

At tree level, the mass of the lightest $CP$-even eigenstate $h$ is bounded, $m_h\leq m_Z$. As emphasized before,
if the soft supersymmetry-breaking mass scale is $M_S\simeq 1$ TeV, radiative corrections can lift the bound to
$m_h\lesssim 130$ GeV. In the limit of large $m_A$ (the decoupling limit, in which $h$ has SM-like couplings) and keeping only contributions proportional to $\alpha_3$ and the top Yukawa, the two-loop analytic formula for $m_h$ reads~\cite{Carena:1995bx}
\begin{eqnarray}
m_h^2 &=& m_Z^2\cos^22\beta\left(1-\frac{3}{8\pi^2}\frac{m_t^2}{v^2}t\right)\nonumber\\
&+&\frac{3}{4\pi^2}\frac{m_t^4}{v^2}\left[\frac{1}{2}X_t+t+\frac{1}{16\pi^2}\left(\frac{3}{2}\frac{m_t^2}{v^2}-32\pi\alpha_3\right)\left(X_tt+t^2\right)\right],
\end{eqnarray}
where
\begin{equation}
\label{Xt}
X_t=\frac{2a_t^2}{M_S^2}\left(1-\frac{a_t^2}{12M_S^2}\right)
\end{equation}
and
\begin{equation}
t=\log\frac{M_S^2}{m_t^2}.
\end{equation}

In Eq.~(\ref{Xt}), $a_t$ is the stop squark mixing parameter given by $a_t\equiv A_t-\mu/\tan\beta$, where $A_t$ is the soft trilinear stop parameter and $\mu$ is the Higgsino mass parameter. The running parameters $\alpha_3$ and $m_t$ are evaluated at the scale of the top quark mass. From this formula it is clear that the upper bound on $m_h$ is achieved for large $\tan\beta$ in the ``maximal mixing" scenario, where $a_t=\sqrt{6}M_S$.

The mixing angle $\alpha$ satisfies the tree level relation
\begin{eqnarray}
\frac{m_A^2+m_Z^2}{m_A^2-m_Z^2}\cdot(\cot\alpha-\tan\alpha) = (\cot\beta-\tan\beta),
\label{alpha}
\end{eqnarray}
where $-\frac{\pi}{2}\leq\alpha\leq 0$. Assuming CP conservation, the couplings of the $CP$-even Higgs $h$ and $H$ to gauge
bosons are given by the Standard Model couplings multiplied by $\sin(\beta-\alpha)$ and
$\cos(\beta-\alpha)$, respectively. The tree level couplings of $h$ and $H$ to down-type fermions are the Standard Model
Yukawa couplings multiplied by $-\sin\alpha/\cos\beta$ and $\cos\alpha/\cos\beta$ respectively, and the couplings
to up-type fermions are similarly rescaled by $\cos\alpha/\sin\beta$ and $\sin\alpha/\sin\beta$.\footnote{For large values of $\tan\beta$, the down-type quark couplings are affected by loop corrections that may be of $\mathcal{O}(1)$~\cite{deltamb1} and alter the production cross sections and branching ratios of the non-standard Higgs bosons in a significant way~\cite{Carena:1998gk},\cite{Coarasa:1995yg}.}
This implies that for associated production of $h$ with a $W$ boson,
\begin{equation}
\frac{\sigma_{MSSM}}{\sigma_{SM}}=\sin^2(\beta-\alpha),
\end{equation}
whereas for the production of $h$ via gluon fusion through a loop of top quarks,
\begin{equation}
\frac{\sigma_{MSSM}}{\sigma_{SM}}\approx\frac{\cos^2\alpha}{\sin^2\beta}.
\end{equation}
The latter ratio is only valid in the limit of heavy superpartners, because it ignores
the possible contribution of squark loops. In our analysis we will instead use the fact that
the gluon fusion cross section is directly proportional to the Higgs decay to $gg$:
\begin{equation}
\frac{\sigma_{MSSM}}{\sigma_{SM}}=\frac{\Gamma_{MSSM}(h\rightarrow gg)}{\Gamma_{SM}(h\rightarrow gg)}.
\end{equation}

\section{Constraints and the $R$ Parameter in the Standard Model}

\begin{figure}[ht]
\begin{center}
\resizebox{120mm}{!}{\includegraphics[width=0.45\textwidth]{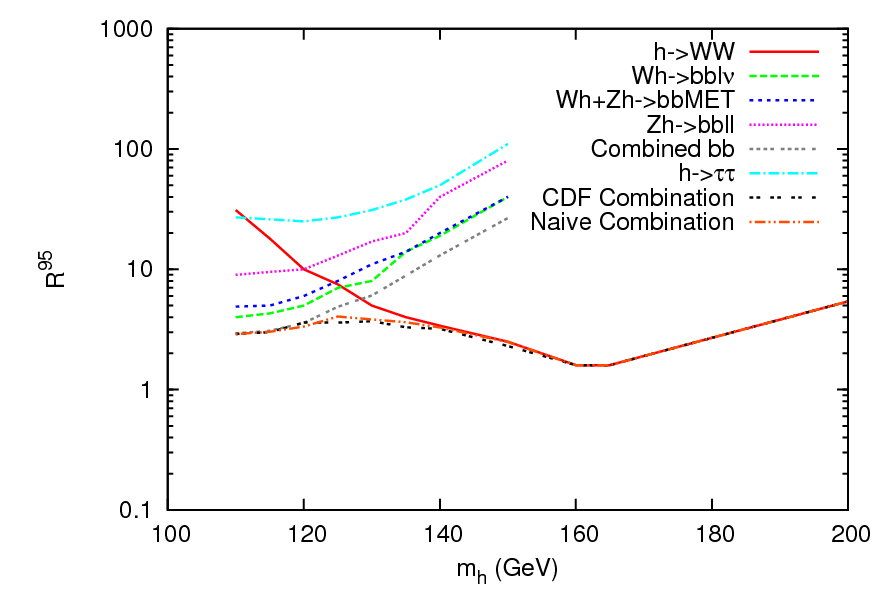}}
\caption{Individual channel and combined constraints on $R_{SM}$ at 95\% C.L. from CDF.
The orange curve gives the na\"{\i}ve analytic combination.}
\label{figcombCDF}
\end{center}
\end{figure}

\begin{figure}[ht]
\begin{center}
\resizebox{120mm}{!}{\includegraphics[width=0.45\textwidth]{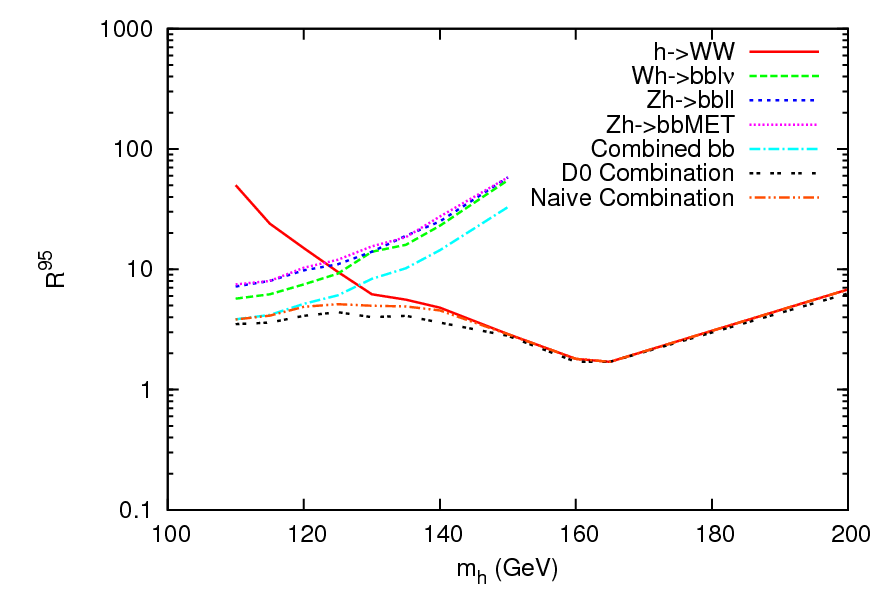}}
\caption{Individual channel and combined constraints on $R_{SM}$ at 95\% C.L. from D$\emptyset$. The orange curve gives the na\"{\i}ve analytic combination.}
\label{figcombD0}
\end{center}
\end{figure}

\begin{figure}[ht]
\begin{center}
\resizebox{120mm}{!}{\includegraphics[width=0.45\textwidth]{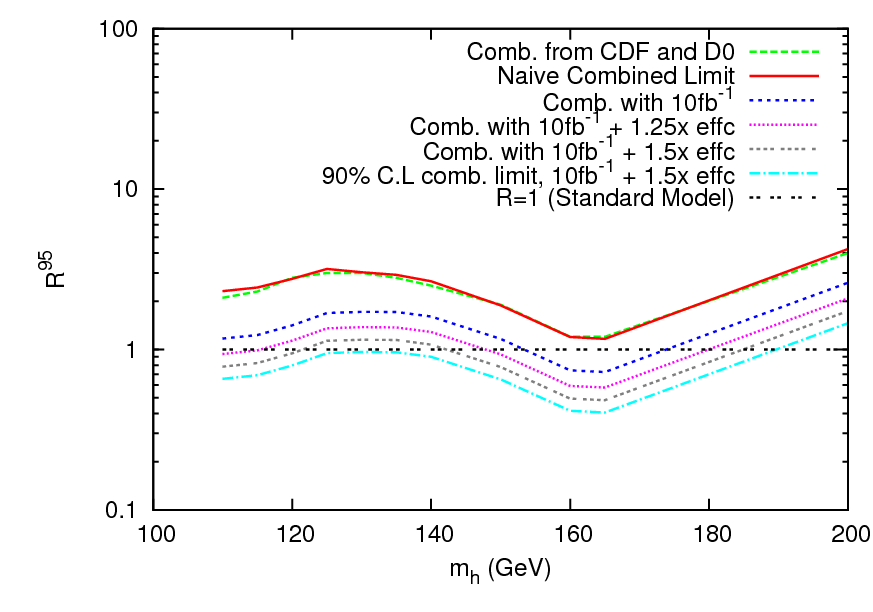}}
\caption{Combined constraints on $R_{SM}$ at 95\% C.L. from CDF, D$\emptyset$, and the combination of the two. Also presented are projected limits after increasing the luminosity to 10 fb$^{-1}$ and including 25-50\% efficiency improvements.}
\label{figcombCDFD0}
\end{center}
\end{figure}

The plots produced by CDF and D$\emptyset$ (\cite{CDFSM},\cite{D0SM}) provide upper limits at 95\% C.L. on the parameter $R_{SM,i}$,
which is the Higgs signal in the decay channel $i$, normalized to the signal in
that channel predicted by the Standard Model:
\begin{equation}
R_{SM,i}\equiv\frac{\sigma_{i}Br_{i}}{\sigma_{SM,i}Br_{SM,i}}.
\end{equation}
Na\"{\i}ve estimates for the limits are obtained by assuming no signal is seen and the errors are only statistical:
\begin{equation}
\label{1}
R^{95}_{SM,i}\equiv 1.96\times\frac{\sqrt{b_i}}{s_{SM,i}},
\end{equation}
where $s_{SM,i}$ and $b_i$ are the expected signal and background from the SM in the
$i$th channel. An upper limit derived from the combination of all channels is also provided by the experimental collaborations~\cite{CDFSM},\cite{D0comb}. $R_{SM,comb}$ is defined to be the Higgs signal in all channels normalized to the predicted SM signal, and therefore parametrizes a space
of theories that differ from the SM only by a universal rescaling of all production cross
sections. Its 95\% C.L. upper limit $R^{95}_{SM,comb}$ can be reproduced to good approximation
by combining Gaussian errors from each channel:
\begin{eqnarray}
\label{2}
R^{95}_{SM,comb}&=&1.96\times\frac{\sqrt{b_{comb}}}{s_{SM,comb}}\nonumber\\
&=&1.96\times\frac{1}{\sqrt{\sum_i\frac{(s_{SM,i})^2}{b_i}}}\nonumber\\
&=&\frac{1}{\sqrt{\sum_i\frac{1}{(R^{95}_{SM,i})^2}}}.
\end{eqnarray}
To illustrate the effectiveness of this approximation, in Figs.~\ref{figcombCDF}
and ~\ref{figcombD0} we show the $R^{95}_{SM}$ curves from individual channels and the
combined constraints as presented by CDF and D$\emptyset$. We include the following search channels: $h\rightarrow b\bar{b}$ with the Higgs produced in association with $W$ or $Z$ bosons, the $h\rightarrow\tau^+\tau^-$ inclusive search, and the $gg\rightarrow h\rightarrow W^+W^-$ search. For each experiment we overlay the combined curve obtained from Eq.~(\ref{2}), and find that it is in good agreement with
the combined result presented by each collaboration. In Fig.~\ref{figcombCDFD0}, we combine
the results from CDF and D$\emptyset$, again comparing with the overall combined curve given
by the collaborations~\cite{Phenomena:2009pt}. Note that with these curves, the mass range $160-170$ GeV is not excluded.
This is because we are combining the \emph{expected} limits from each channel, rather than
the experimentally observed limits. A downward fluctuation was observed in the analysis of the
$h\rightarrow W^+W^-$ channel, leading to the exclusion. The reason we take the expected limits is so
that we can rescale them by projected increases in luminosity and signal efficiency. The overall
SM Higgs limits corresponding to three such improvement scenarios are also presented in
Fig.~\ref{figcombCDFD0}. The first assumes an increase in luminosity up to 10~fb$^{-1}$ per experiment
in all channels. Such an increase in luminosity is expected, if the Tevatron continues running
until mid-2011. The second assumes the same luminosity increase, as well as 25\% efficiency improvements
in the detection of the $b\bar{b}$ and $W^+W^-$ Higgs decay channels. The final scenario increases the efficiency improvements further to 50\%. It is important to note that each of these efficiency improvements scale down the previous upper bound by about 20\%. This is approximately the same reduction achieved by increasing the luminosity from 7~fb$^{-1}$ to 10~fb$^{-1}$, or by decreasing the statistical significance from 95\% to 90\%. Therefore, our results can be reinterpreted in this context: if the luminosity only reaches 7~fb$^{-1}$, an efficiency
improvement of 50\% in the $b\bar{b}$ and $W^+W^-$ channels will be necessary to reach the exclusion limit achieved with 10~fb$^{-1}$ and a 25\% increase in efficiency. If only 25\% is gained in efficiency and 7~fb$^{-1}$ in luminosity, the same exclusions will apply at 90\% C.L.

We emphasize that our na\"{\i}ve combination leads to a conservative upper limit on the $R$ parameter, up to about 10\% larger than that presented by the experimental collaborations in some ranges of Higgs mass. Furthermore, it does not include the possible effects of new search channels. For example, one channel not included in our analysis is the $h\rightarrow\gamma\gamma$ from D$\emptyset$, which, from the results of Ref.~\cite{gammagamma}, would improve the na\"{\i}ve combined bound on $R$ by up to 1.5\% in the $120-140$ GeV mass range. Therefore, our improvement factors imply upper bounds on those required for large coverage of the SM Higgs mass range.


\begin{figure}[ht]
\begin{center}
\resizebox{120mm}{!}{\includegraphics[width=0.45\textwidth]{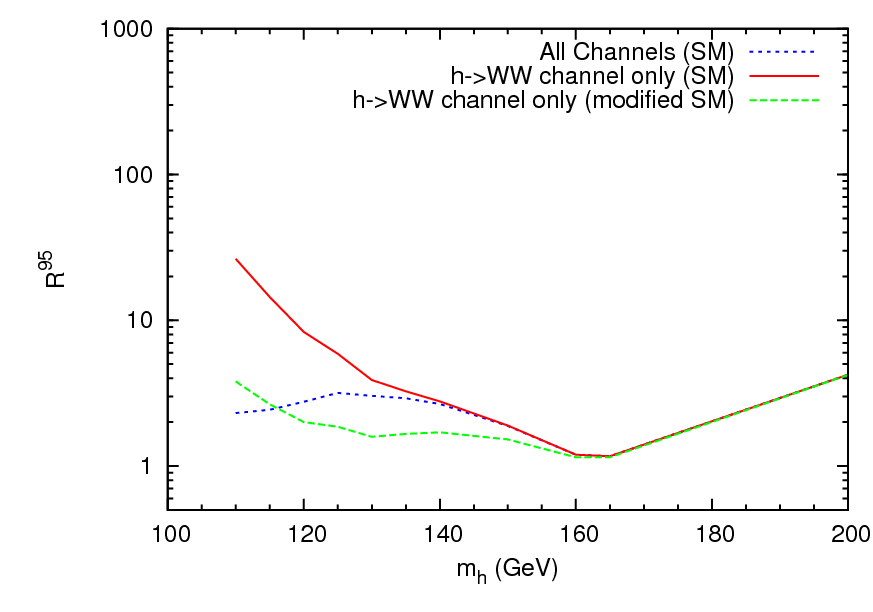}}
\caption{The bound on $R$ at 95\% C.L. from the $h\rightarrow W^+W^-$ channel in the SM and in a modified SM
scenario where the coupling of $h$ to down-type fermions is suppressed. The combined constraint from
all channels is shown for reference.}
\label{fignobb}
\end{center}
\end{figure}

It is clear from Figs.~\ref{figcombCDF}-~\ref{figcombCDFD0} that Higgs masses near $125$ GeV will be
the most difficult to probe. It is essential in this region to include constraints from both $b\bar{b}$
and $W^+W^-$ decay channels. As a test of the strength of the $W^+W^-$ limit in the low mass region, in Fig.~\ref{fignobb}
we plot the limit in the SM and in a modification of the SM where the Higgs coupling to down-type fermions
is highly suppressed, leading to an enhancement of the branching ratio to $W^+W^-$. This scenario arises in a
certain window of MSSM parameters. We can see from Fig.~\ref{fignobb} that the bound is quite strong
($R^{95}\lesssim 2$ for $120$GeV$\lesssim m_h\lesssim 170$GeV) even without any improvements in
efficiency or luminosity.

\section{The $R$ Parameter in the MSSM}

The combined constraint cannot be used immediately to place limits on the MSSM, because as
mentioned previously it is a constraint on models that differ from the SM by a single constant
rescaling of $\sigma\times Br$ for all channels. In the MSSM, different channels will be
related to the SM channels by different rescalings. Therefore, each channel should be individually
scaled so that the $R^{95}$ constraint bounds the Higgs signal normalized to the expected signal from the MSSM:

\begin{equation}
\label{3}
R^{95}_{MSSM,i}=R^{95}_{SM,i}\times\frac{\sigma_{SM,i}Br_{SM,i}}{\sigma_{MSSM,i}Br_{MSSM,i}}.
\end{equation}

Then, the $R^{95}_{MSSM,i}$ can be combined as in Eq.~(\ref{2}) above to give a net constraint
on the MSSM.

We include both the light and the heavy $CP$-even MSSM Higgs bosons as intermediate states for each channel.
The most sensitive Tevatron searches for SM-like Higgs bosons have been in the $b\bar{b}$ channel with associated production of the Higgs, and the $W^+W^-$ channel with Higgs production via gluon fusion. These channels require sizeable Higgs gauge couplings, and therefore have been primarily
relevant for the light Higgs $h$, which has SM-like gauge couplings in most of parameter space.
However, as $m_A$ decreases the mixing angle becomes larger and $m_H$ becomes smaller,
so $H$ becomes increasingly SM-like and consequently plays a bigger role in the constraints
inferred from these searches.

In contrast to $b\bar{b}$, the $\tau^+\tau^-$ channel search is inclusive~\cite{tautau}, and therefore conservative limits can be inferred for both SM-like and nonstandard Higgs searches. As an upper limit on the associated production channel, it can be interpreted as a constraint from an SM-like search. This constraint is typically quite weak compared to the limit from the $b\bar{b}$ channels, as can be seen from Fig.~\ref{figcombCDF} and assuming the rescaling factor in Eq.~(\ref{3}) is $\mathcal{O}(1)$. In the following we will always lump this SM-like $\tau^+\tau^-$ constraint together with the $b\bar{b}$ constraint. On the other hand, when the $\tau^+\tau^-$ data is taken as a limit on the gluon fusion production channel, the constraint from the CP-odd and nonstandard CP-even Higgs bosons can be quite strong~\cite{Abulencia:2005kq},\cite{:2008hu}. These particles have $\tan^2\beta$ enhanced production rates through loops of bottom quarks, and so the rescaling factor relative to the SM can be significant if they are sufficiently light. In the following, when we refer to the $\tau^+\tau^-$ constraint, we mean this constraint coming from the nonstandard Higgs search.

Our strategy will be as follows: we pick benchmark scenarios for all the MSSM parameters except
for $\tan\beta$ and $m_A$, which are the dominant parameters affecting the Higgs signal.
We scan over the $(m_A, \tan\beta)$ plane, calculating the spectrum and the scaling factors
$\sigma_{SM,i}Br_{SM,i}/(\sigma_{MSSM,i}Br_{MSSM,i})$ for all channels. The masses and
branching ratios are computed numerically using HDECAY \cite{Djouadi:1997yw}, and in particular
the numerator is calculated at the Standard Model Higgs mass equal to the mass of the $CP$-even
MSSM Higgs in the intermediate state (we checked that similar results are obtained by using
CPsuperH~\cite{CPsuperH}). Finally we read off the expected $R^{95}_{SM,i}$ from the
CDF and D$\emptyset$ plots and use Eqs.~(\ref{3}) and (\ref{2}) to compute the value of $R^{95}$
at each point in the parameter space.

As emphasized before, we will first present our results for the constraints from the SM-like Higgs search channels
and the $gg\rightarrow h,H\rightarrow\tau^+\tau^-$ nonstandard search channel separately. This will demonstrate the capabilities of the separate searches in covering the MSSM parameter space. At the end we will
combine the constraints to see the complementarity of the coverage.

\section{Benchmark Scenarios}

We consider four benchmark sets of soft parameters \cite{Carena:2002qg}, which enter in
the dominant loop corrections to the Higgs mass matrix. All soft sfermion masses are set to a common value $M_S$ and the top quark mass is set to the current central value of $173.1$ GeV~\cite{:2009ec}. We scan over $m_A$ from $100$ GeV to $300$ GeV in 100 steps, and $\tan\beta$ from $3$ to $60$ also in 100 steps. The first benchmark point is the case of maximal mixing,
with
\begin{eqnarray*}
M_S=1 \mbox{ TeV, }& &a_t=\sqrt{6}M_S\mbox{,}\nonumber\\
\mu=200 \mbox{ GeV, }& &M_2=200\mbox{ GeV,}\nonumber\\
A_b=A_t\mbox{, }& &m_{\tilde{g}}=0.8M_S.
\end{eqnarray*}
As mentioned previously, this choice of parameters leads to the largest radiative addition
to the lightest Higgs mass.
The second point is the opposite scenario of minimal stop mixing,
\begin{eqnarray*}
M_S=2 \mbox{ TeV, }& &a_t=0\mbox{,}\nonumber\\
\mu=200 \mbox{ GeV, }& &M_2=200\mbox{ GeV,}\nonumber\\
A_b=A_t\mbox{, }& &m_{\tilde{g}}=0.8M_S.
\end{eqnarray*}
Here the mass of the lightest Higgs is lower, $m_h\sim 113-118$ GeV, and the constraint
will be stronger, as can be seen from the Standard Model combined curve. The soft supersymmetry-breaking mass scale
is increased in this scenario in order to avoid LEP bounds~\cite{LHWG:2005}. We chose a value of $\mu$ of order $m_Z$. A larger value of $\mu$, of the order of the stop masses, would lead to a decrease in the lightest CP-even Higgs mass of about 2~GeV in most of the parameter space, without affecting the quality of Higgs searches in the analyzed channels.

The third benchmark lies in the ``gluophobic" region, where the gluon fusion mechanism for production
of the SM-like Higgs is suppressed by a cancellation between the top quark loop and the stop
loop~\cite{Djouadi:1998az}. This requires a light stop\footnote{For an analysis of Tevatron Higgs searches for an alternative MSSM scenario associated with electroweak baryogenesis (in which a light stop enhances gluon fusion), see~\cite{Menon:2009mz}.}, which can be achieved for a smaller
scale $M_S$ and moderate, negative values of the parameter $a_t$ controlling the stop mixing. For this point, we take
\begin{eqnarray*}
M_S=350 \mbox{ GeV, }& &a_t=-770\mbox{ GeV,}\nonumber\\
\mu=300 \mbox{ GeV, }& &M_2=300\mbox{ GeV,}\nonumber\\
A_b=A_t\mbox{,  }& &m_{\tilde{g}}=500\mbox{ GeV}.
\end{eqnarray*}

The final point is in the ``small $\alpha_{eff}$" scenario, where $\mu$ and $A_t$ take typical
intermediate magnitudes but opposite signs. In this case the branching fraction of $h\rightarrow b\bar{b}$
becomes highly suppressed for a certain range of $m_A$ and $\tan\beta$, due to a cancellation in the
off-diagonal component of the neutral CP-even Higgs boson mass matrix between tree-level terms
suppressed by $1/\tan\beta$ and loop corrections suppressed by factors of $1/(16 \pi^2)$. Without an
off-diagonal term in the mass matrix, the lightest $CP$-even Higgs becomes entirely up-type,
and does not couple to bottom quarks at tree level (a small coupling appears at the
loop-level~\cite{Carena:1998gk}).
The cancellation can be seen explicitly from the loop-corrected off-diagonal matrix element,

\begin{eqnarray}
\mathcal{M}_{12}^2&\simeq&-\left[m_A^2+m_Z^2-\frac{h_t^4v^2}{8\pi^2}(3\bar{\mu}^2-\bar{\mu}^2\bar{A_t}^2)\right]\sin\beta\cos\beta\\
&+&\left[\frac{h_t^4v^2}{16\pi^2}\bar{\mu}\tilde{a_t}(\bar{A_t}\tilde{a_t}-6)\sin^2\beta+\frac{3h_t^2m_Z^2}{32\pi^2}\bar{\mu}\tilde{a_t}\right]\left[1+\frac{t}{16\pi^2}(4.5h_t^2-0.5h_b^2-16g_3^2)\right].\nonumber
\end{eqnarray}

\begin{figure}[ht]
\begin{center}
\begin{tabular}{cc}
\resizebox{70mm}{!}{\includegraphics{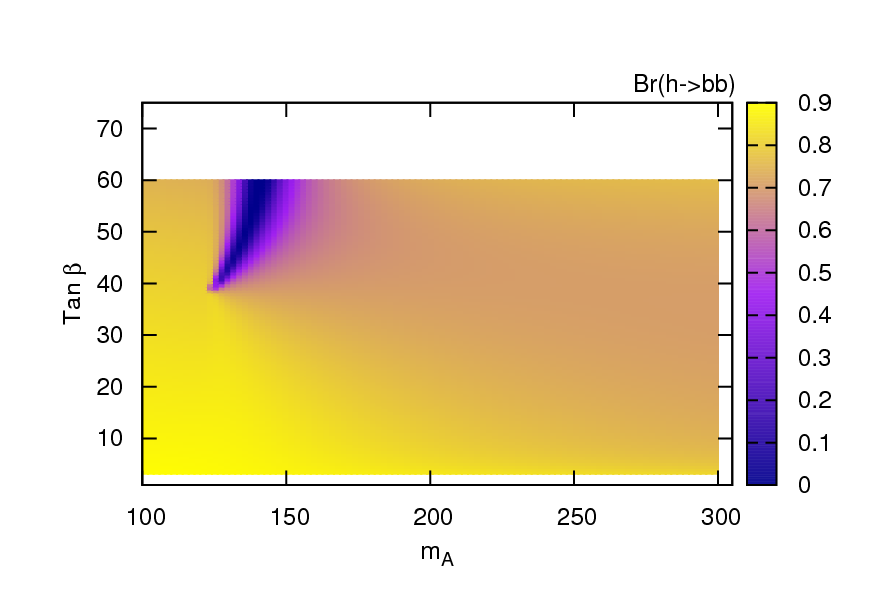}} &
\resizebox{70mm}{!}{\includegraphics{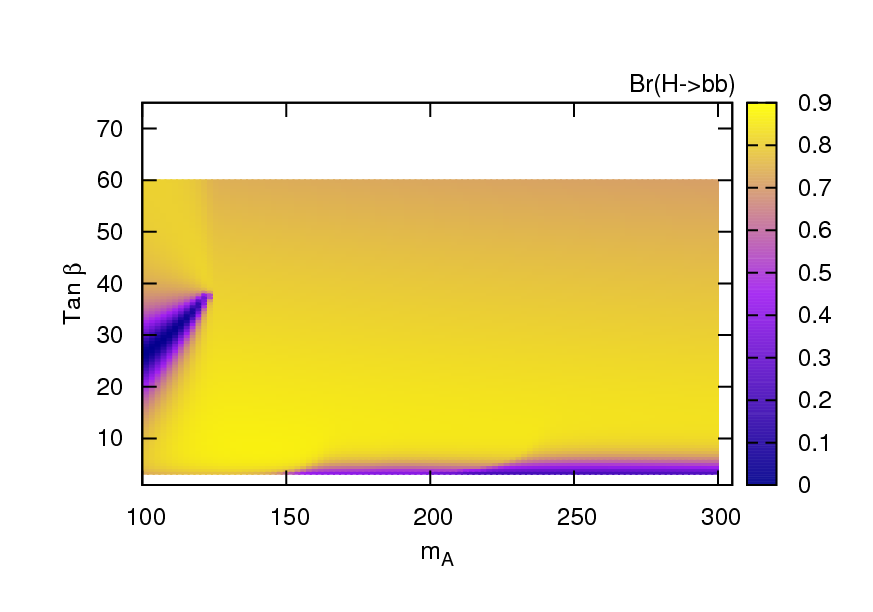}} \\
\end{tabular}
\caption{(1)Br($h\rightarrow b\bar{b}$) and (2) Br($H\rightarrow b\bar{b}$) in the small $\alpha_{eff}$ scenario.}
\label{figbbh}
\end{center}
\end{figure}

\begin{figure}[ht]
\begin{center}
\begin{tabular}{cc}
\resizebox{70mm}{!}{\includegraphics{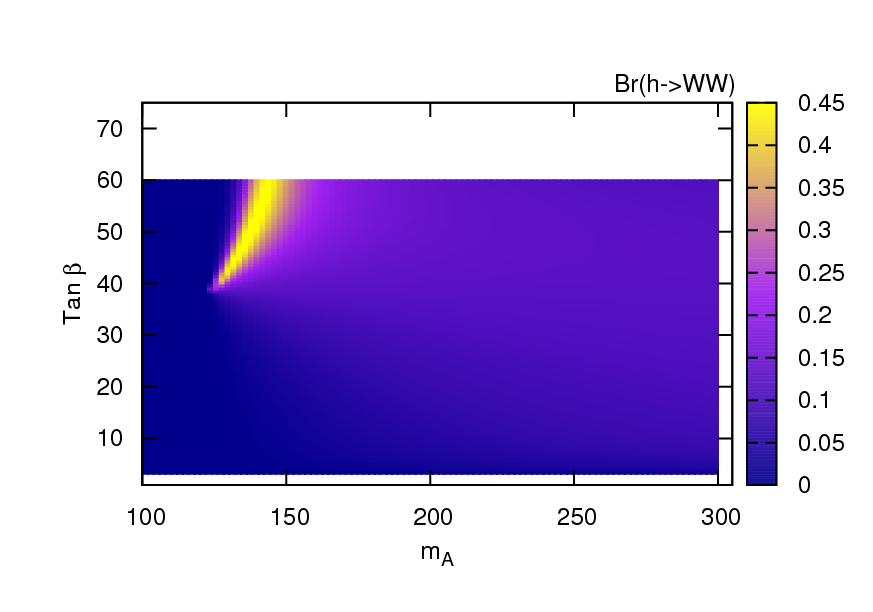}} &
\resizebox{70mm}{!}{\includegraphics{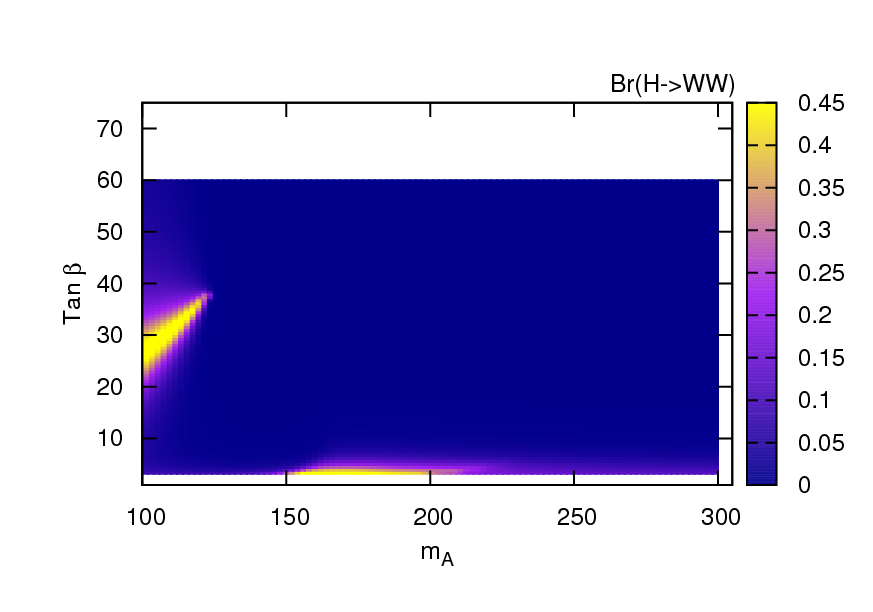}} \\
\end{tabular}
\caption{(1) Br($h\rightarrow W^+W^-$) and (2) Br($H\rightarrow W^+W^-$) in the small $\alpha_{eff}$ scenario.}
\label{figWWh}
\end{center}
\end{figure}

Here $h_t$ and $h_b$ are Yukawa couplings, $g_3$ is the strong running coupling, $\tilde{a_t}=a_t/M_S$, $\bar{\mu}=\mu/M_S$, $\bar{A_t}=A_t/M_S$, and $t=\log(M_S^2/m_t^2)$. For moderate to large $\tan\beta$, $\mathcal{M}_{12}^2$ vanishes if
the following approximate relationship holds:
\begin{equation}
\left[\frac{m_A^2}{m_Z^2}-\frac{1}{2\pi^2}(3\bar{\mu}^2 -\bar{\mu}^2\bar{A_t}^2)+1\right]
\simeq\frac{\tan\beta}{150}\left[\bar{\mu}\bar{A_t}(2\bar{A_t}^2-11)\right].
\end{equation}
In the maximal mixing case, this relation cannot be satisfied, and in the no-mixing case, it can only be
satisfied for very large $\bar{\mu}$. On the other hand, if $\mu=-A_t\sim M_S$ the relation can
be easily satisfied for low $m_A$.  The benchmark values are\footnote{The values of $a_t$ and $\mu$ chosen here differ from those presented in the small-$\alpha_{eff}$ scenario of Ref~.\cite{Carena:2002qg}. The values here allow a greater contribution to the limit from the $WW$ decay modes of the SM-like Higgs in the regions where its $b\bar{b}$ mode is suppressed. All figures including those corresponding to the benchmark values in \cite{Carena:2002qg} are available on the website listed in the introduction.}
\begin{eqnarray*}
M_S=800 \mbox{ GeV, }& &a_t=-1.5\mbox{ TeV,}\nonumber\\
\mu=1.5 \mbox{ TeV, }& &M_2=500\mbox{ GeV,}\nonumber\\
A_b=A_t\mbox{, }& &m_{\tilde{g}}=500\mbox{ GeV}.
\end{eqnarray*}
Fig.~\ref{figbbh} demonstrates the suppression of the $b\bar{b}$ branching fractions
of the light and heavy Higgs bosons, and Fig.~\ref{figWWh} shows the corresponding increase in the branching
fractions to $W^+W^-$. The apex of the cones is the point where the mass eigenvalues become degenerate,
and so the mixing angle and consequently the cross sections and decay widths vary rapidly in the
$(m_A, \tan\beta)$ parameter space.

\section{Projections and Analysis}

Following Eq.~(\ref{1}), $R^{95}$ scales inversely the square root of the luminosity and
inversely with the signal efficiency. We study the improvements necessary for the Tevatron
to exclude regions of MSSM parameter space at 90\% and 95\% C.L. As in the SM case discussed above, we take as milestones
total luminosities of 10 fb$^{-1}$ per experiment in all channels, and improvement factors
of 1.0, 1.25, and 1.5 in the signal efficiency in the $b\bar{b}$ and $W^+W^-$ channels. We assume no efficiency
improvement in the $\tau^+\tau^-$ non-standard Higgs boson searches.

\begin{figure}[ht]
\begin{center}
\resizebox{120mm}{!}{\includegraphics[width=0.45\textwidth]{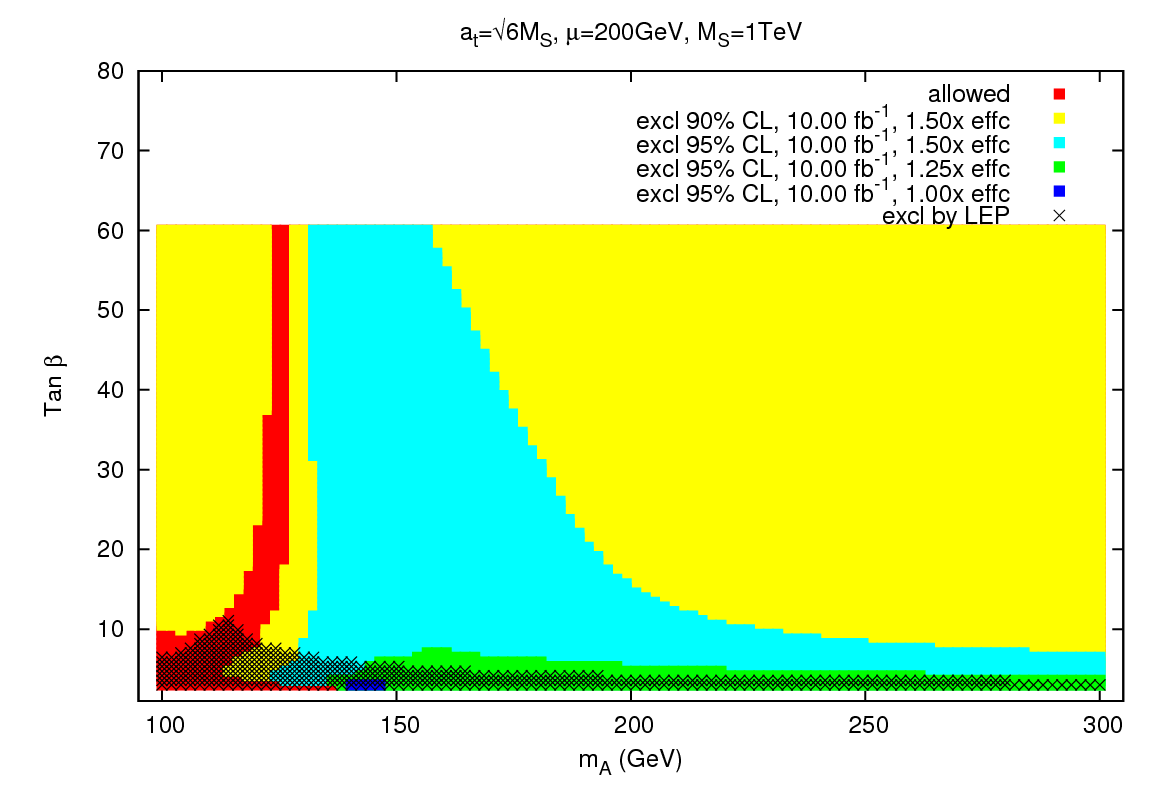}}
\caption{Exclusion limits at 90\% and 95\% C.L. in the maximal mixing scenario of the MSSM, including only $b\bar{b}$ and $W^+W^-$ decay channels. }
\label{Amax}
\end{center}
\end{figure}

Fig.~\ref{Amax} gives the projected limits from the $b\bar{b}$ and $W^+W^-$ channels in the maximal mixing
scenario. For large values of $m_A\gg m_Z$, $H$ decouples, and the couplings of $h$ approach the
Standard Model values. In this regime $h$ is mostly up-type as $\alpha$ becomes small and negative, but the mixing angle
suppression of the coupling to down-type quarks is compensated by the $\tan\beta$ enhancement of the Yukawa.
The lightest CP-even Higgs mass in the decoupling limit is close to 125~GeV, which as observed before,
is the most difficult mass range for SM-like Higgs bosons searches at the Tevatron collider (see Fig.~3).
Hence, large improvements in efficiencies and luminosity will be necessary to probe this scenario.

For moderate, fixed $m_A$ and decreasing $\tan\beta$, the coupling of $h$ to down-type quarks is
enhanced by a more rapid increase in $\sin\alpha$ than in $\cos\beta$. Consequently the constraints
are stronger than in the SM. At lower values of $m_A\gtrsim m_h$, this effect becomes more pronounced,
so a large area can be excluded at 95\% C.L. As $m_A$ becomes equal to or smaller than about $125\mbox{ GeV}$,
however, $\alpha\lesssim -\frac{\pi}{4}$ and the production rate
of $h$ by standard processes, which involve couplings to gauge bosons and to the top quark, is substantially decreased for moderate to large $\tan\beta$.  Fortunately, in this region $m_H$ is
light ($m_H\approx 126\mbox{ GeV}$) and its production rate by standard processes
is growing as the production of $h$ falls. The combination of constraints from $h$ and $H$ still
produces a moderate exclusion limit in a significant region of the parameters.

\begin{figure}[ht]
\begin{center}
\resizebox{120mm}{!}{\includegraphics[width=0.45\textwidth]{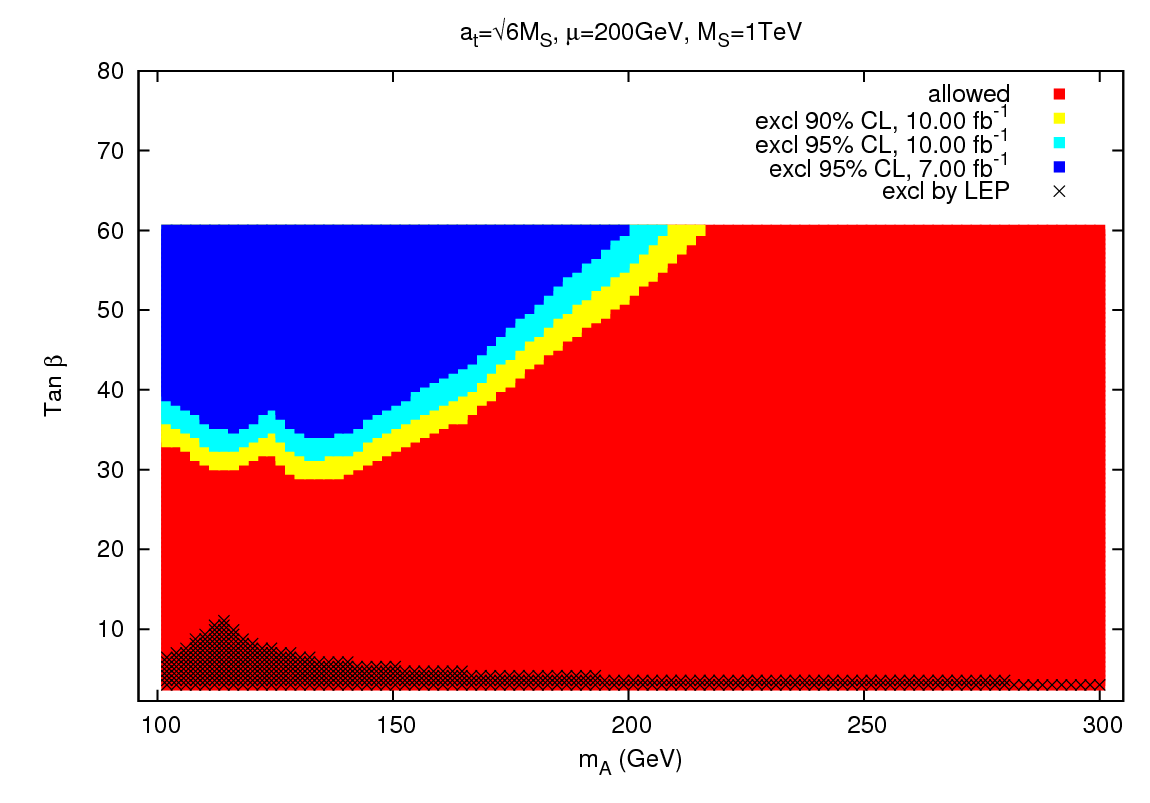}}
\caption{Exclusion limits at 90\% and 95\% C.L. in the maximal mixing scenario of the MSSM, including only the $\tau^+\tau^-$ inclusive search. No efficiency improvements are applied. }
\label{Amaxtonly}
\end{center}
\end{figure}

\begin{figure}[ht]
\begin{center}
\resizebox{120mm}{!}{\includegraphics[width=0.45\textwidth]{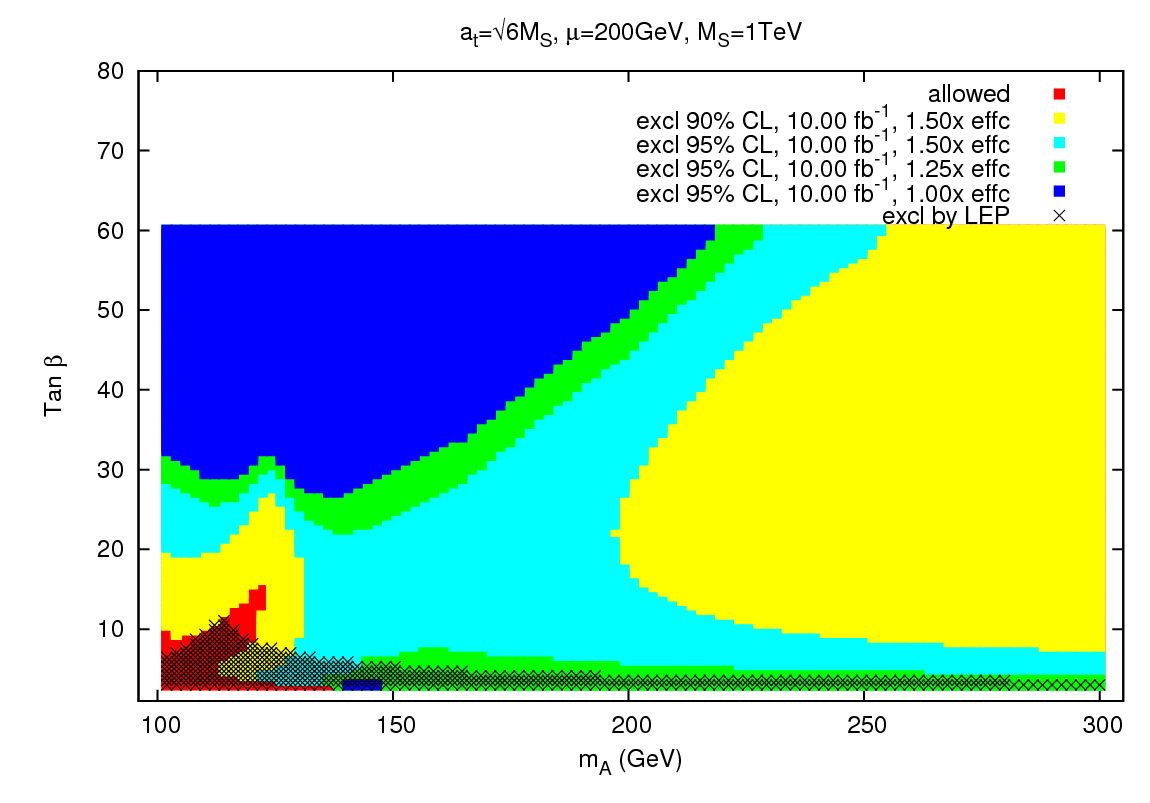}}
\caption{Exclusion limits at 90\% and 95\% C.L. in the maximal mixing scenario of the MSSM, including all channels. }
\label{Amaxwt}
\end{center}
\end{figure}

In Fig.~\ref{Amaxtonly} we present the limits coming only from the $\tau^+\tau^-$ channel search for the
nonstandard Higgs. In the low $m_A$ and large $\tan\beta$  regime, $h$ has suppressed gauge
couplings and is dominantly down-type. It exchanges this role with $H$ as $m_A$ increases towards
the decoupling limit. In either case the nonstandard Higgs is light enough to be produced, mainly
by gluon fusion through $b$ and $\tilde{b}$ loops. Its coupling to down-type fermions is enhanced by
$\tan\beta$, so the constraint can be significant without any improvements in efficiency or luminosity.
The mass and fermionic couplings of the $CP$-odd Higgs $A$ mimic those of the nonstandard Higgs, so we
take $2\times \sigma\times Br$ to approximate its contribution to the $\tau^+\tau^-$ constraint, which,
as has been shown in Ref.~\cite{Carena:2005ek}, is only weakly dependent on the value of the soft
breaking parameters~\cite{deltamb1}.
In the intense coupling regime, $m_A\sim m_h \sim m_H$, both $H$ and $h$ have sizeable down-type components,
leading to the strongest constraints from $\tau^+\tau^-$. Fig.~\ref{Amaxwt} presents the combined limits from both
the SM-like and nonstandard Higgs searches. We see that much of parameter space requires a 50\% signal
efficiency improvement in the $b\bar{b}$ and $W^+W^-$ channels in order to make a test this scenario at
the 90\% C.L..  The decoupling limit, where $h$ is indistinguishable from a 125 GeV SM Higgs, is the most difficult
to probe.

\begin{figure}[ht]
\begin{center}
\resizebox{120mm}{!}{\includegraphics[width=0.45\textwidth]{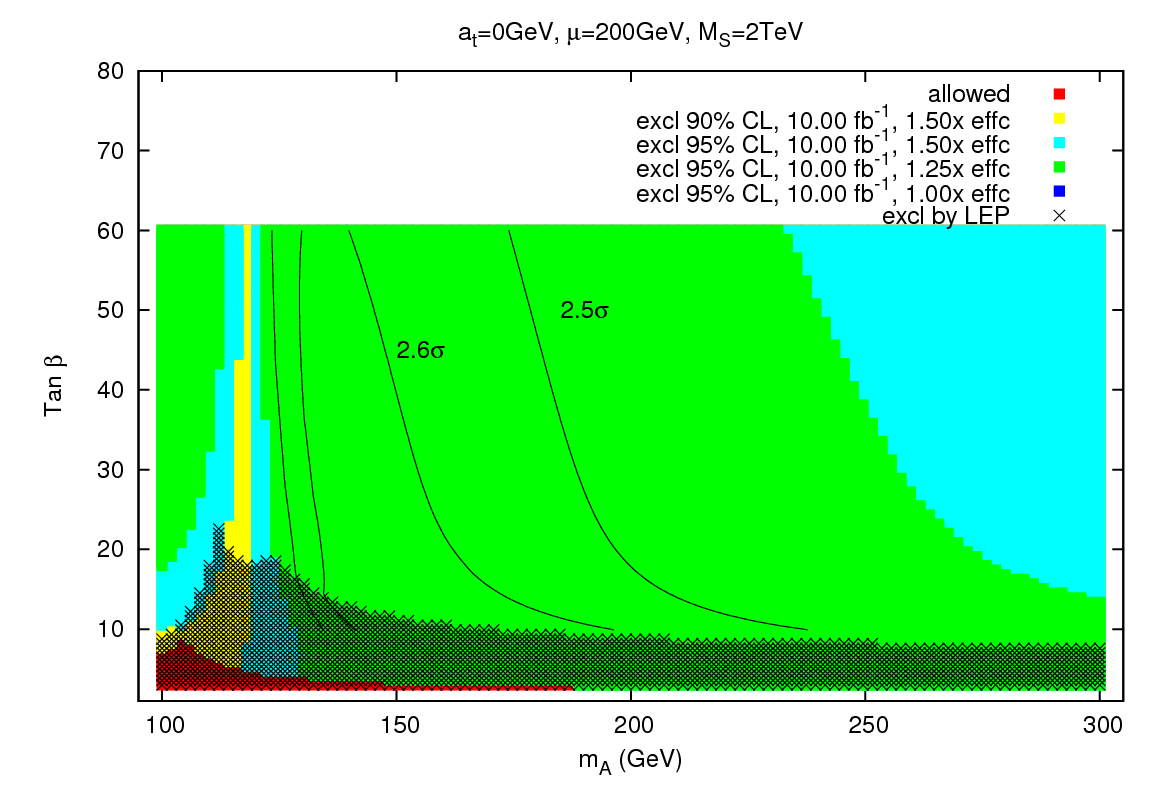}}
\caption{Exclusion limits at 90\% and 95\% C.L. in the no-mixing scenario of the MSSM, including only $b\bar{b}$ and $W^+W^-$ decay channels. }
\label{Amin}
\end{center}
\end{figure}

\begin{figure}[ht]
\begin{center}
\resizebox{120mm}{!}{\includegraphics[width=0.45\textwidth]{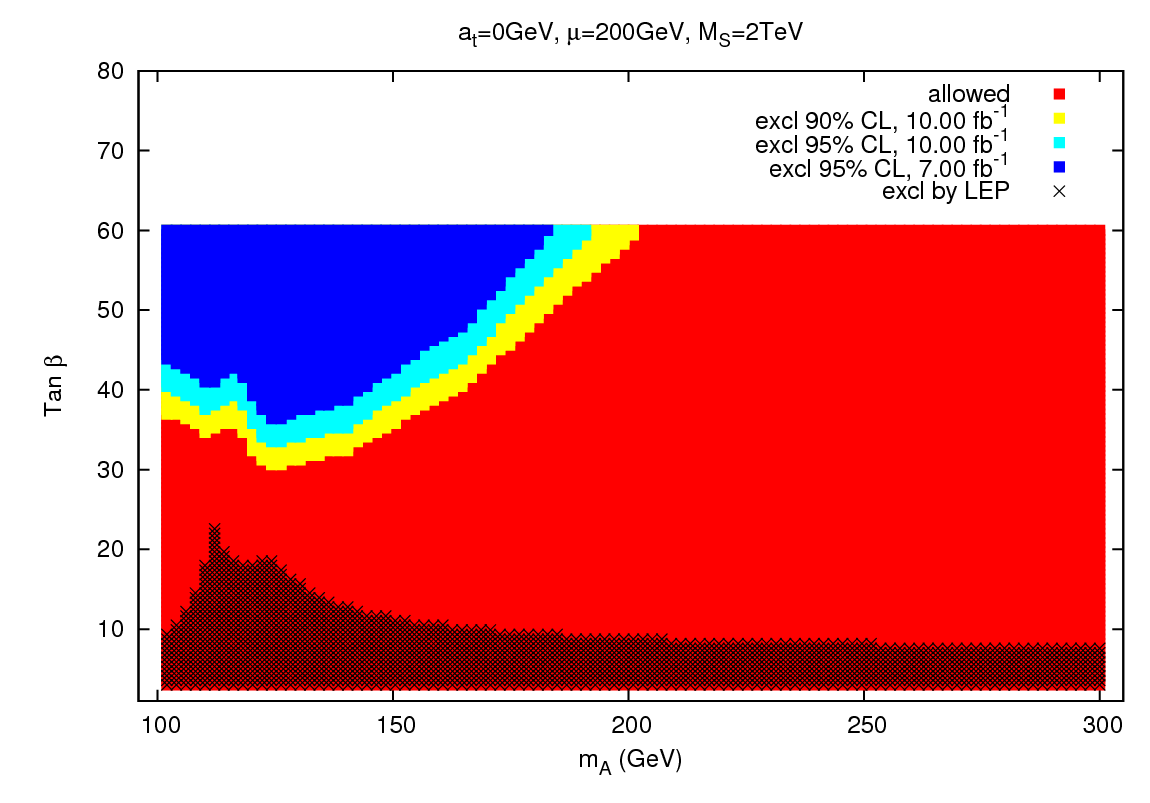}}
\caption{Exclusion limits at 90\% and 95\% C.L. in the no-mixing scenario of the MSSM, including only the $\tau^+\tau^-$ inclusive search. No efficiency improvements are applied. }
\label{Amintonly}
\end{center}
\end{figure}

\begin{figure}[ht]
\begin{center}
\resizebox{120mm}{!}{\includegraphics[width=0.45\textwidth]{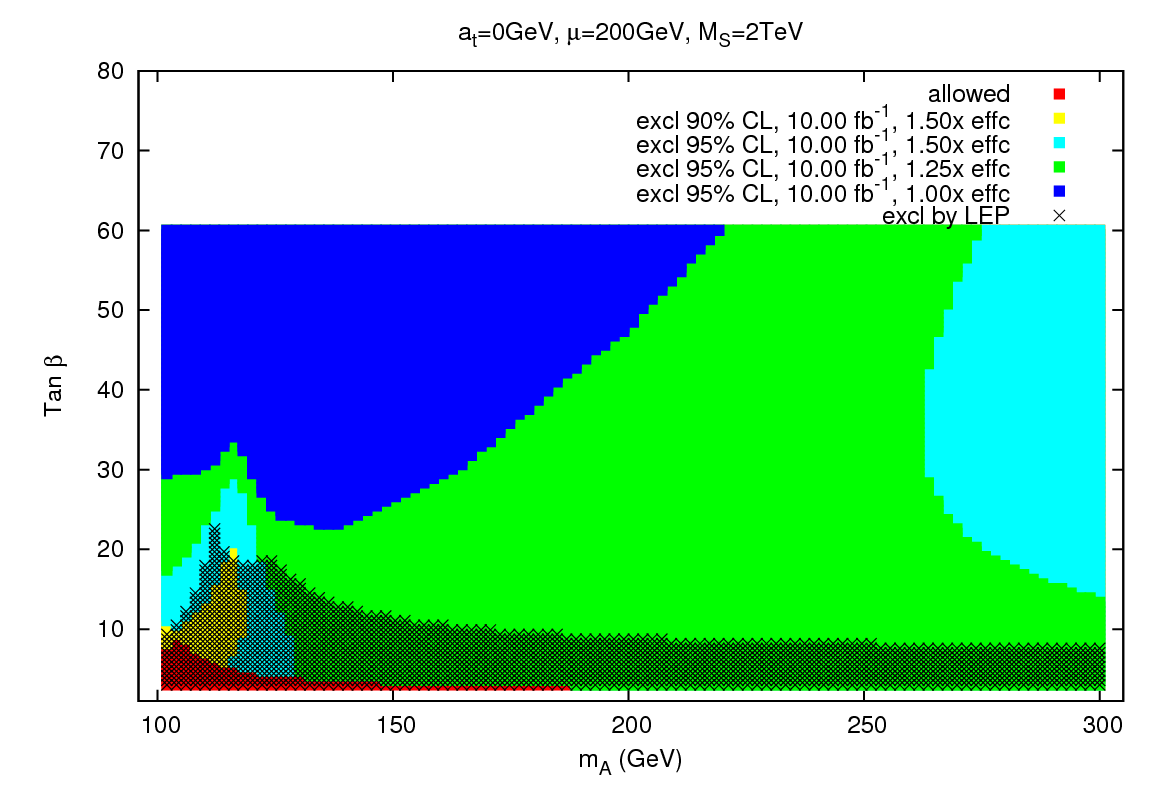}}
\caption{Exclusion limits at 90\% and 95\% C.L. in the no-mixing scenario of the MSSM, including all channels. }
\label{Aminwt}
\end{center}
\end{figure}

Fig.~\ref{Amin} gives the projected limits from the $b\bar{b}$ and $W^+W^-$ channels in the no-mixing scenario.
The behavior of the production and couplings is similar, but smaller loop corrections from the squark
sector reduce $m_h$ to around $117$~GeV in most of parameter space. These small Higgs boson masses are characteristic of
constrained models of supersymmetry, like the CMSSM, in which the $A_t$ parameter is rarely larger
than the characteristic stop masses.
The constraints are stronger than in the maximal mixing case
due to the decrease of $R^{95}_{SM}$ (for a previous analysis in the CMSSM, based on somewhat more
optimistic projections of the reach in the $b\bar{b}$ channel, see, for example,
Ref.~\cite{Roszkowski:2006mi}). The black contours in Fig.~\ref{Amin} show the regions of maximal statistical significance that can be achieved in SM-like searches with the full 50\% efficiency improvements. Fig.~\ref{Amintonly} gives the constraints from only the $\tau^+\tau^-$ channel,
and Fig.~\ref{Aminwt} gives the combined limit. We see that all of parameter space can be covered
at 95\% C.L. if the total integrated luminosity increases to 10~fb$^{-1}$ and
there are 25\% efficiency improvements in the SM-Higgs search channels.

\begin{figure}[ht]
\begin{center}
\resizebox{120mm}{!}{\includegraphics[width=0.45\textwidth]{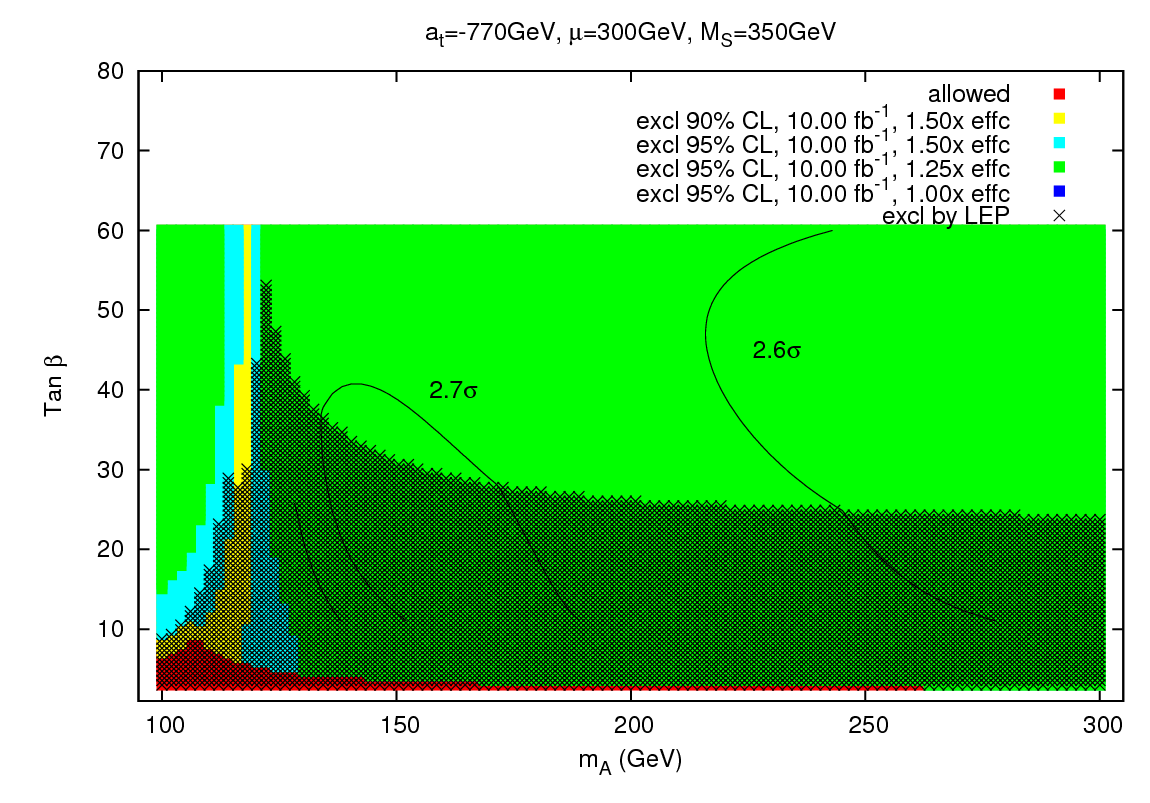}}
\caption{Exclusion limits at 90\% and 95\% C.L. in the gluophobic scenario of the MSSM, including only $b\bar{b}$ and $W^+W^-$ decay channels. }
\label{Aglue}
\end{center}
\end{figure}

\begin{figure}[ht]
\begin{center}
\resizebox{120mm}{!}{\includegraphics[width=0.45\textwidth]{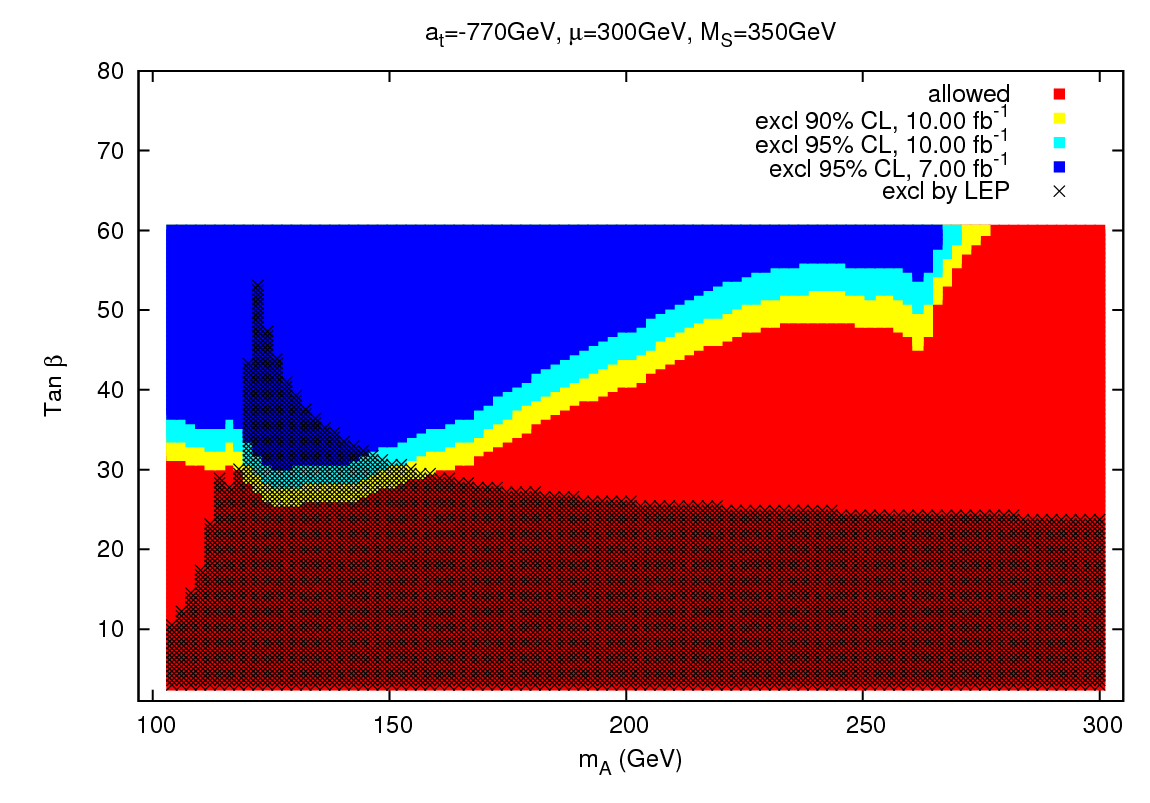}}
\caption{Exclusion limits at 90\% and 95\% C.L. in the gluophobic scenario of the MSSM, including only the $\tau^+\tau^-$ inclusive search. No efficiency improvements are applied.}
\label{Agluetonly}
\end{center}
\end{figure}

\begin{figure}[ht]
\begin{center}
\resizebox{120mm}{!}{\includegraphics[width=0.45\textwidth]{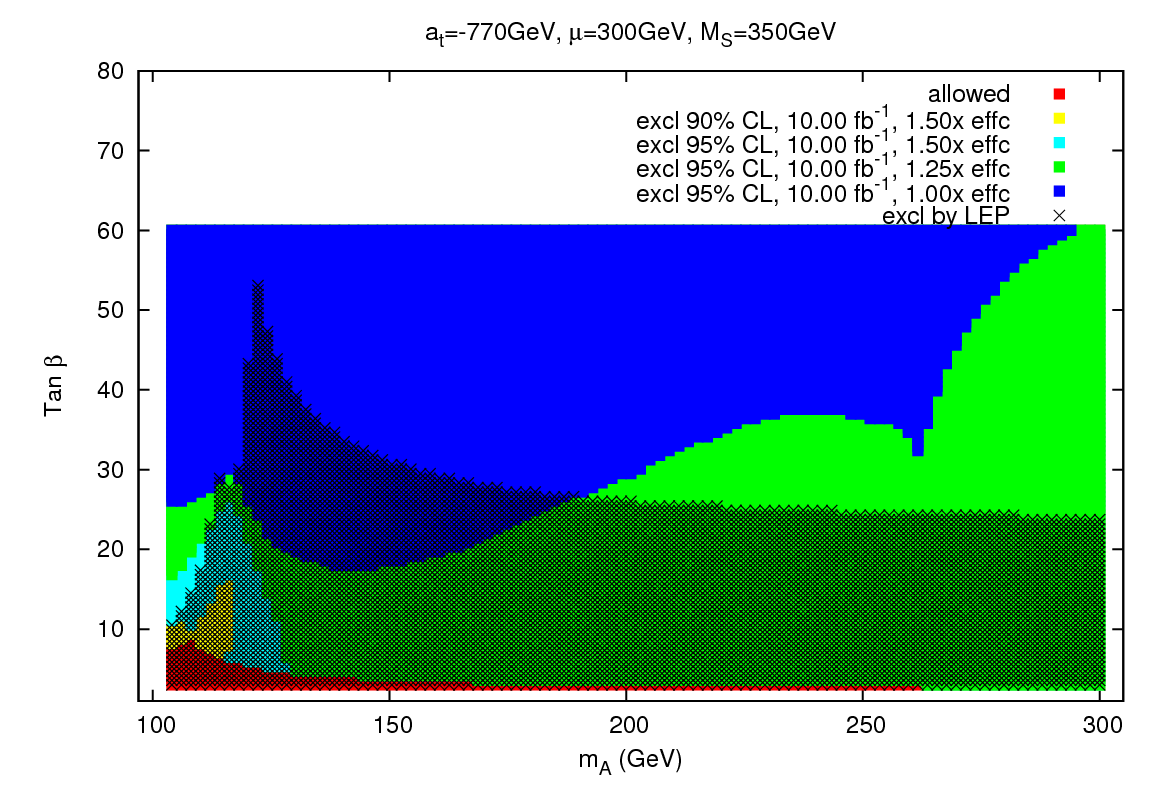}}
\caption{Exclusion limits at 90\% and 95\% C.L. in the gluophobic scenario of the MSSM, including all channels. }
\label{Agluewt}
\end{center}
\end{figure}

Fig.~\ref{Aglue} gives the projected limits from the $b\bar{b}$ and $W^+W^-$ channels in the gluophobic region.
The suppression of gluon fusion spoils the utility of the $h\rightarrow W^+W^-$ channel, but the SM-like
Higgs mass is lower and so the $b\bar{b}$ constraints are dominant. This leads to a result qualitatively
very similar to what we found in the no-mixing scenario. Fig.~\ref{Agluetonly} gives the limits from just
the $\tau^+\tau^-$ channel, which are again similar to the no-mixing case because gluon fusion production of the
nonstandard Higgs bosons, which is governed by down-type quark and squark loops, is not suppressed. There is a feature at $m_A\simeq260$ GeV which is caused by the opening of the $H\rightarrow \tilde{t_1}\tilde{t}^*_1$ decay channel, where $m_{\tilde{t}_1}\simeq130$ GeV. Just below threshold, the light stop loop enhances the gluon fusion cross section. Above threshold, the rapid rise of the branching fraction of $H\rightarrow \tilde{t_1}\tilde{t}^*_1$ sharply suppresses the $H\rightarrow \tau^+\tau^-$ decay mode. Fig.~\ref{Agluewt} gives the combined limit.

\begin{figure}[ht]
\begin{center}
\resizebox{120mm}{!}{\includegraphics[width=0.45\textwidth]{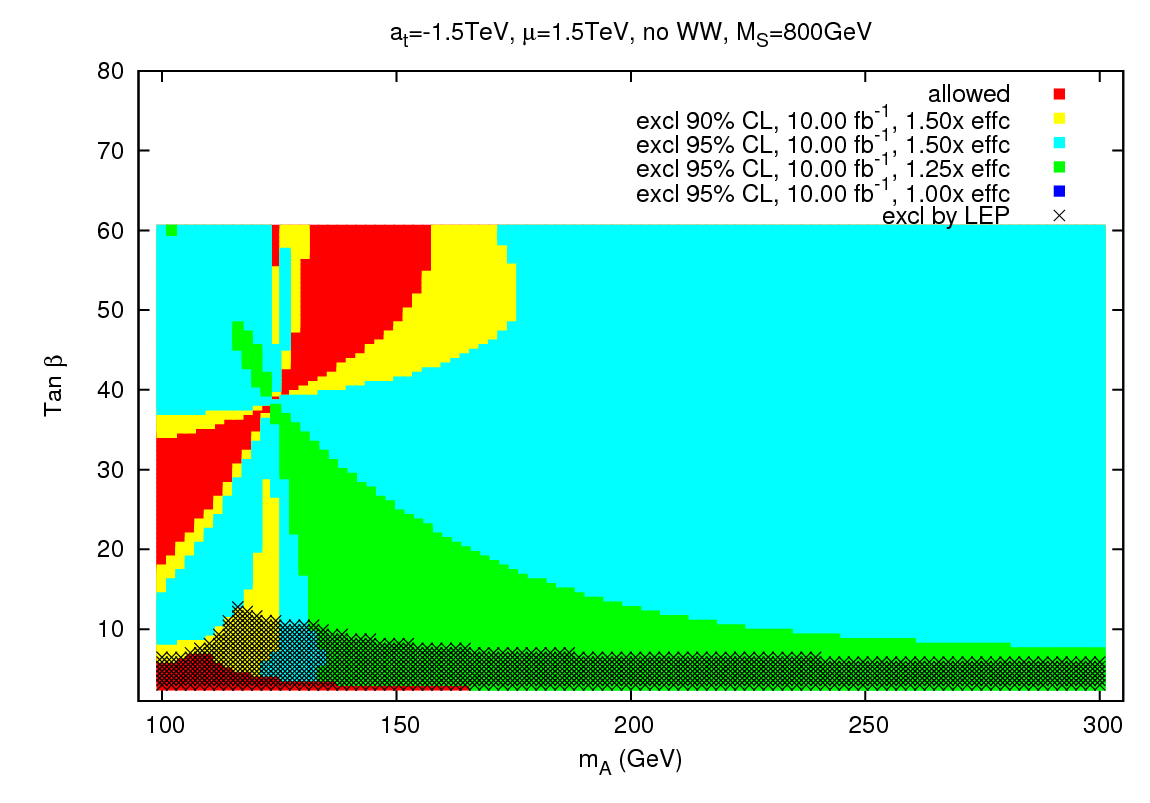}}
\caption{Exclusion limits from the $h,H\rightarrow b\bar{b}$ channels at 90\% and 95\% C.L. in the small $\alpha_{eff}$ scenario of the MSSM. }
\label{Afree1}
\end{center}
\end{figure}

\begin{figure}[ht]
\begin{center}
\resizebox{120mm}{!}{\includegraphics[width=0.45\textwidth]{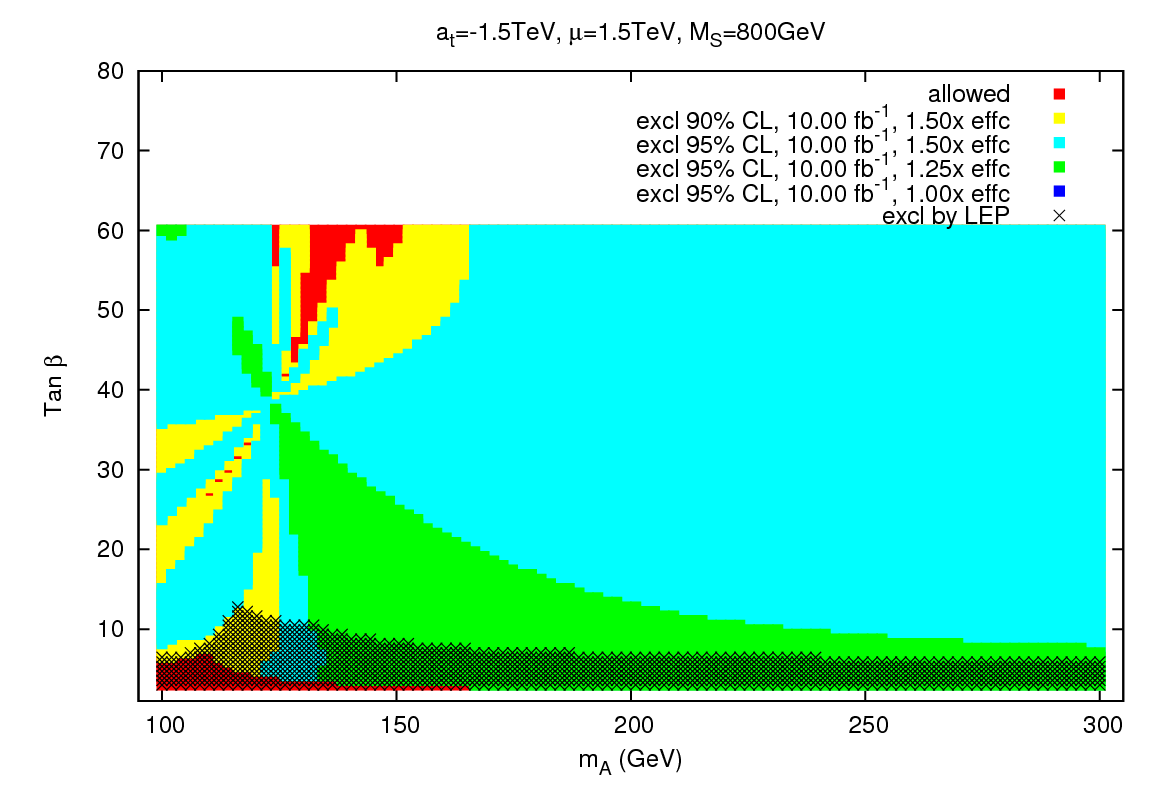}}
\caption{Exclusion limits from the $h,H\rightarrow b\bar{b}$ and $h,H\rightarrow W^+W^-$ channels at 90\% and 95\% C.L. in the small $\alpha_{eff}$ scenario of the MSSM. }
\label{Afree2}
\end{center}
\end{figure}

\begin{figure}[ht]
\begin{center}
\resizebox{120mm}{!}{\includegraphics[width=0.45\textwidth]{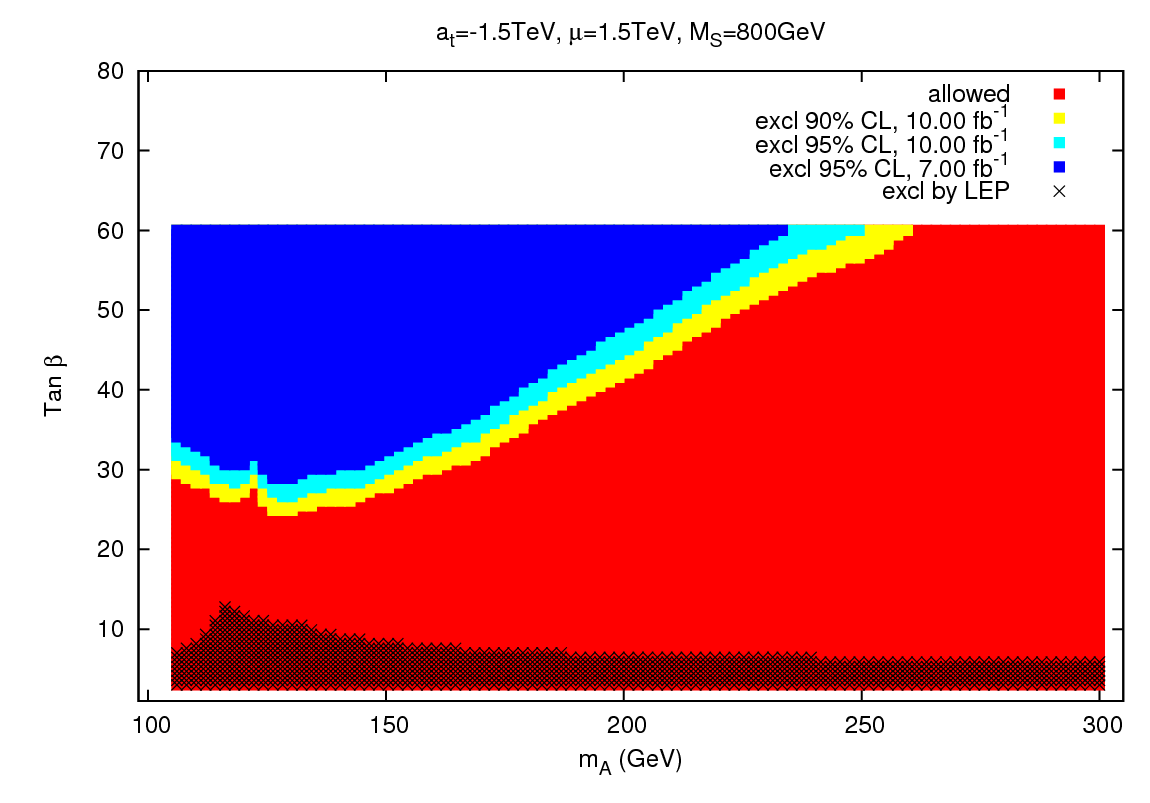}}
\caption{Exclusion limits at 90\% and 95\% C.L. in the small $\alpha_{eff}$ scenario of the MSSM, including only the $\tau^+\tau^-$ inclusive search. No efficiency improvements are applied. }
\label{Afreetonly}
\end{center}
\end{figure}

\begin{figure}[ht]
\begin{center}
\resizebox{120mm}{!}{\includegraphics[width=0.45\textwidth]{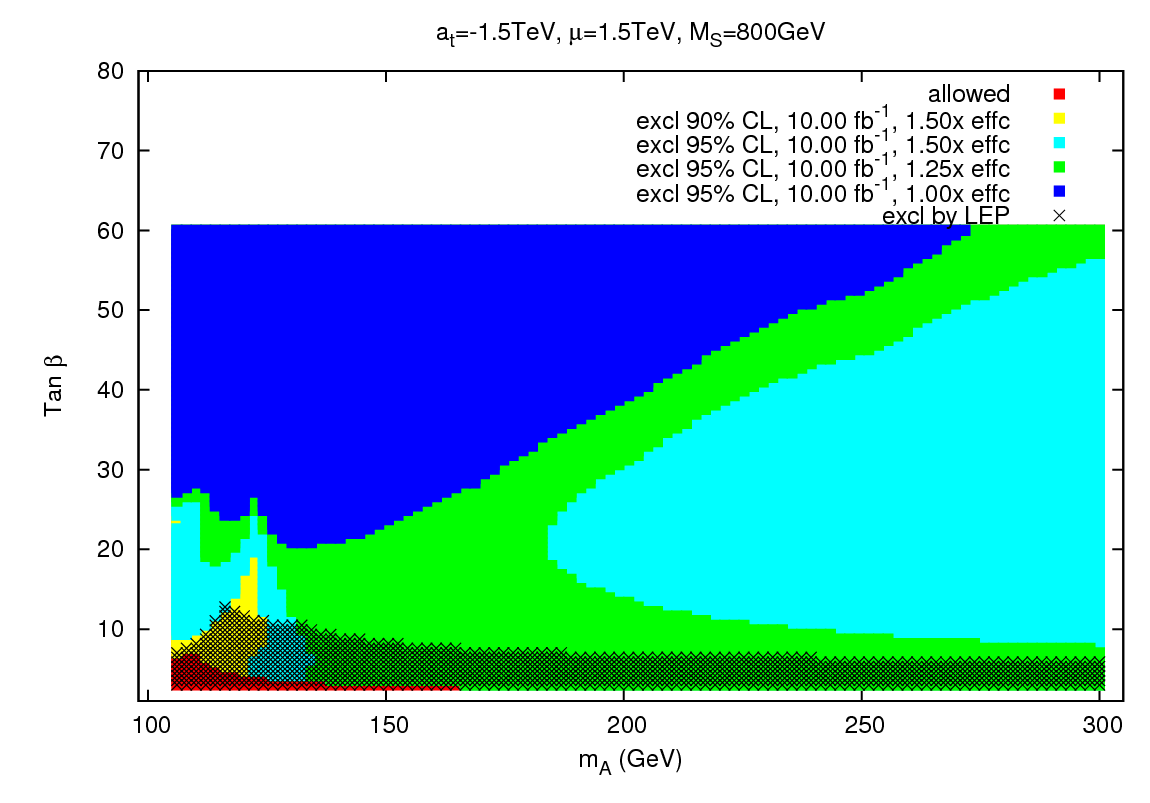}}
\caption{Exclusion limits at 90\% and 95\% C.L. in the small $\alpha_{eff}$ scenario of the MSSM, including all channels. }
\label{Afreewt}
\end{center}
\end{figure}

 Finally, we present the projected limits from SM-like Higgs searches in the small
$\alpha_{eff}$ scenario.
In order to show the importance of the $h,H \to W^+W^-$ Higgs decay channels, in
Fig.~\ref{Afree1} we present the results omitting the contributions from
these channels. There is a well-known stripe of parameter space unprobed by the $b\bar{b}$ searches because of the
loop-induced cancellation of $\mathcal{M}_{12}^2$~\cite{Carena:1998gk}. In Fig.~\ref{Afree2} we
present the same results, but now including
the constraint from the $h,H\rightarrow W^+W^-$ channels. We find that the $W^+W^-$ channel can cover almost
all of this previously inaccessible window at 90\% C.L., with sufficient improvements.
In Fig.~\ref{Afreetonly} we give the limit from the $\tau^+\tau^-$ channel alone. Although it covers the
region unprobed by the $b\bar{b}$ channels, it is no longer so crucial for covering all of the $(m_A,\tan\beta)$
plane, because of the limit from the $W^+W^-$ channel. Fig.~\ref{Afreewt} demonstrates the complementarity
of the searches.


\section{Conclusions}

In this note we have studied the improvements necessary for the Tevatron to probe the Higgs sector of the minimal
supersymmetric extension of the SM. If the experiments can achieve the increases in
luminosity and signal efficiency studied in this work, the Tevatron may be able to probe significant regions
of the MSSM parameter space to 95\% C.L., and probe all of the parameter space at 90\% C.L.

In particular, if 10~fb$^{-1}$ of integrated luminosity are achieved, a 25\% increase in efficiency of the $b\bar{b}$ channel (or a similar improvement coming from the addition of other, complementary channels) will be enough to probe scenarios with small values of $A_t$ at 95\% C.L. If only 7~fb$^{-1}$ of integrated luminosity are gathered, a 50\% increase in efficiency is needed to probe these scenarios at the same level. Similar results were found in the gluophobic scenario, since the necessary light stops push the Higgs mass to low values without affecting the $b\bar{b}$ production channel in a significant way. On the other hand, if $A_t$ acquires a larger value, close to that which maximizes the SM-like Higgs mass, both an increase in luminosity to about 10~fb$^{-1}$ and of efficiencies by a factor~1.5 will be necessary to fully probe the MSSM Higgs sector. In addition, non-standard Higgs searches in the inclusive $\tau^+ \tau^-$ channel provide valuable complemetarity to the SM-like searches in order to cover the small $m_A$, large $\tan\beta$ region of this scenario. Finally, the complementarity of SM-like searches in the $b\bar{b}$ and $W^+W^-$ channels was shown to be important in scenarios in which the $h b \bar{b}$ coupling is suppressed.


\section*{Acknowledgments}

Work at ANL is supported in part by the U.S. Department of Energy (DOE), Div.~of HEP, Contract DE-AC02-06CH11357.
Work at EFI is supported in part by the DOE through Grant No. DE-FG02- 90ER40560.  T.L. is also supported by the
Fermi-McCormick Fellowship. This work was supported in part by the DOE under Task TeV of contract DE-FGO2-96-ER40956.

\end{document}